# Performing clinical drug trials in children with a rare disease

*Victoria Hedley, Rebecca Leary, Anando Sen, Anna Irvin, Emma Heslop and Volker Straub*

John Walton Muscular Dystrophy Research Centre, Newcastle University and Newcastle Hospitals NHS Foundation Trust, Newcastle upon Tyne, United Kingdom

## What conditions are necessary to achieve clinical trial readiness?

### The burden of rare and pediatric diseases

A large proportion of pediatric diseases are also classed as rare diseases: "rare disease" meaning a condition affecting no more than 5 per 10,000 people, according to the European definition set-down in 2000 by Regulation (EC) No 141/2000 (EU, 2000). Between 6 and 8000 separate conditions fall into this definition of an RD. These are largely genetic, multisystem conditions, and unsurprisingly, the conditions are very heterogenous: some will sadly result in death very soon after birth or become severely disabling early in childhood. In fact the majority of rare diseases affect children, collectively, they are responsible for 35% of deaths in the first year of life and are a significant cause of pediatric hospital admissions; one-third of children born with a rare disease will not live to see their fifth birthday (Rode, 2005, p. 5; Yoon et al., 1997). Other rare diseases only manifest in later life and do not reduce lifespan but can pose other problems. Several features, however, tend to be common to all rare diseases. For instance, the rarity of any single condition means not only a small number of directly affected patients, by definition, but also a scarcity of expertise: it is unlikely that patients and families will live close to an expert in their condition. Patients with a rare disease often face what is known as a diagnostic odyssey (Anna Kole & Faurisson, 2009), waiting many years for a correct diagnosis: the pursuit of an accurate diagnosis is hampered by the lack of visibility around where the experts in a particular disease happen to sit, but problems start earlier in the health pathways, with General Practitioners largely (and quite naturally) having limited awareness (McMullan et al., 2021; Walkowiak & Domaradzki, 2021) of rare disease as a concept and traditionally struggling to refer patients to the right centers most able to get them an accurate diagnosis. Rare diseases are poorly understood at most levels of the health and social care services, in all countries. And that boundary between

"care" and "research" is typically more blurred than is the case for common pediatric diseases—the main reason for this is quite simply the absence of robust treatments for pediatric rare diseases (PRDs). In addition, there remain significant challenges in the conduct of pediatric clinical trials including trial planning, innovative methodologies for smaller populations, delays in trial start up and recruitment, recruitment issues due to small and scarce populations, and lack of endpoints (European Medicines Agency and its Paediatric Committee, 2016). Approximately, 94% of all conditions classed as rare have no dedicated treatment. R&D has tended to focus on a relatively limited number of RD—for example, in Europe, at the end of 2018, almost a third (34%) of orphan drug approvals were in the field of oncology: metabolic, endocrine, and hematological disorders accounted for a further 26% leaving many fields neglected therapeutically, by comparison (Tambuyzer et al., 2020).

This naturally creates huge inequities for patients and families and marks out pediatric rare disease as an area of major unmet need. To tackle this, concerted efforts are required, at the national, European, and global levels.

Building optimal policy frameworks for pediatric research to thrive

Orphan and pediatric legislation obviously have a significant impact of the R&D landscape for PRDs (EU, 2000, 2004): Companies need assurance of a sufficient return on investment to engage in this space, and incentives offered by regulations, such as periods of marketing exclusivity, are crucial. The current orphan and pediatric legislative frameworks at EU level are under revision, a process which is generating great interest (Aartsma-Rus et al., 2021), particularly in light of a decreasing European competitiveness in the R&D space broadly. However, beyond all this, significant change is required to speed-up the rate of therapy development (Hivert et al., 2021) and bring more transformative treatments to patients. There is a consensus that a seismic shift in traditional R&D is needed, and for this to occur, policymakers need to make significant changes. Because care and research are so closely connected in PRD, wide-ranging policies are needed. The bulk of this chapter concerns the operational delivery of clinical trials in PRDs; however, many elements need to be considered to bring a PRD to the point at which that community may be deemed "trial-ready," and thousands of conditions are not yet at this stage. Supportive policies are therefore required.

A recent foresight study for Rare Disease, Rare 2030, provided comprehensive recommendations (Kole & Hedley, 2021) to guide the rare disease community toward the scenario deemed most desirable looking ahead to the year 2030 and beyond. This foresight study had a very broad focus precisely because it is difficult to separate policies and recommendations required to enhance, say diagnosis, from those needed to improve clinical trials. The project consortium embraced a classical foresight methodology, first clarifying the European status quo across all major topics under the RD umbrella. Next, a large panel of over 200 experts from 38 countries, encompassing all stakeholder groups, brainstormed on trends noted in recent years, and predicted to continue and/or emerge in the coming decade or two (Kole, 2019). These future-facing trends were clustered, and extrapolated to give four very different futures (Fig. 19.1).

The most popular of these (following a long period of consultation through surveys, conferences, young citizens' groups, and of course the Rare 2030 Panel of Experts itself) was the scenario of "Investment for Social Justice." This is a scenario in which collective accountability triumphs over individualism, which can manifest in various ways. For instance, at population levels, individuals with a PRD are valued in society at large, which is sensitive to and supportive of them accessing adequate diagnostics, treatment, care, and social support. In such a society, those without a rare disease do not regard the high price of some treatments, such a gene therapies, as disproportionate to the lives saved. By another extension, patients with different conditions show solidarity to one another, avoiding the competitive element that can sometimes be noted. As no future will have infinite resources, this *may* mean that disease areas which traditionally have seen the most R&D investment would settle for a slightly slower pace of continued progress, at least initially, to enable RD communities with essentially nothing to share resources. Solidarity also applies to the most desirable model for policymakingdin this preferred scenario, countries reject the siloed approach to policy-making and accelerate collaboration across borders for RD, showing a willingness to learn good lessons from other countries' experiences. This top-ranked scenario is also a future in which patient needs drive innovation (in everything from R&D to care delivery), as opposed to purely market forces.

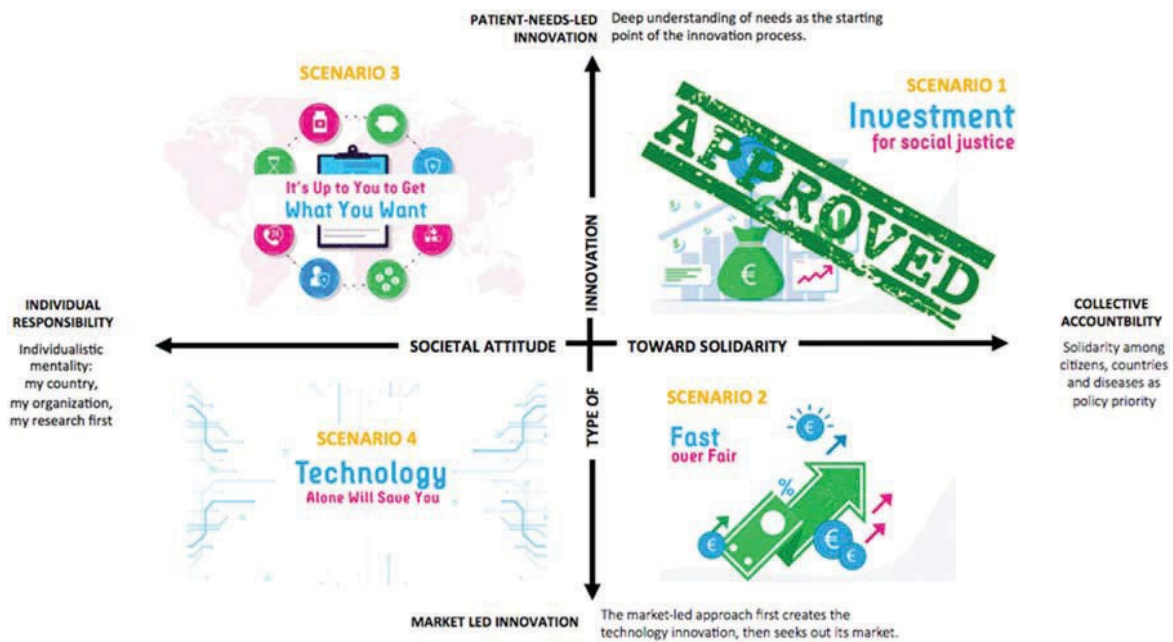

FIGURE 19.1 The four scenarios envisioned by rare 2030. *The image is reproduced from Kole, A., Hedley, V. (2021). Recommendations from the rare 2030 foresight study. The future of rare diseases starts today.* https://download2.eurordis.org/rare2030/Rare2030_recommendations.pdf *with permission.*

To support Europe in reaching this preferred scenario, Rare 2030 published an extensive list of recommendations. They are clustered around large, ambitious goals, and headline messages, with each accompanied by more practical and specific activities to accelerate progress in the following major topics: policy; data; diagnostics; access to care; research; integrated and social care; accessibility and availability of orphan medicines and devices; and patient partnerships. Some of these recommendations are particularly relevant to planning and delivering better clinical trials in PRD, as below. But as many conditions need to be met for clinical trials to take place at all, many of the Rare 2030 recommendations are also very relevant for national and European policymaking, to help build that cohesive ecosystem in which data exploitation and health and social system design conspire to generate knowledge and facilitate patient access to expertise. A holistic and far-reaching strategy to achieving trial-readiness for the thousands of so-called neglected PRDs currently lacking clinical trials should therefore also target the following dimensions.

## Prerequisites for clinical research─understanding of pathogenesis and natural history

To develop or identify robust therapeutic options, it is necessary to isolate the cause of the disease and understand the disease-causing mechanisms. Basic research is still lacking in many PRDs (Zhu et al., 2021). For these reasons, the Rare 2030 foresight study called for actions including the following.

- Resources should be designated to foster research and development in very rare and disregarded conditions which lack therapeutic options.
- More investments, prioritization, and incentives should be ensured for basic and clinical research in areas where these are lacking─research funders must address the significant gap in basic research and discovery science for rare diseases and simultaneously build more bridges to translate innovative and promising research from bench to the clinic and back.
- Investments into public private partnerships operating in the precompetitive space should be increased, with greater coordination and collaboration between funding sources and across sectors, and with particular attention to tech-intensive and other advanced approaches (A Kole & Hedley, 2021)

Building on the last point, the Moonshot for Rare Diseases, announced late in 2022, will advance public‒private partnerships, to cast new light on the neglected rare diseases (many of which will affect the pediatric population) (Rare Disease Moonshot, 2023).

Another reason some PRD remain far from the clinical trial phase is that their natural history remains poorly understood (which tends not to be the case with more common pediatric diseases) (Taruscio et al., 2011). An important source of natural history data for rare diseases, historically, has been registries (Bladen et al., 2013; Boulanger et al., 2020; Gliklich et al., 2014; Mistry et al., 2022).

Registries collect structured data on a given population, and may be established by disease/disease area, to collect data on all patients with a given condition in a given region, nation, continent, or even globally. Alternatively, it may be that registries are set-up at a much broader level, for example, seeking to capture all cases of rare disease within a given geographical area (Domenica Taruscio et al., 2015). Providing that registry datasets are sufficiently broad to capture information on all the main ways in which a condition can impact a patient and/or have been created with a reasonable level of knowledge of these things, registry data collected at intervals on patients with a given disease can yield invaluable insights to the course of a condition, for instance illustrating at which age patients may begin to lose ambulation, develop visual impairment, require ventilation support, etc (Hedley et al., 2019). Because PRD populations are, by definition, limited in size, it has long been accepted that obtaining robust and reliable data by which to identify natural history trends entails pooling data (either via a centralized database or through a more federated data model) across borders, to try to reach a so-called critical mass of patients (Julkowska et al., 2017): for instance, if 90% of 500 patients enrolled in a European-wide registry for condition X develop shortness of breath by age 18, clinicians can more confidently predict this as a likely symptom of the natural course of condition X, than would be the case if 90% of 5 patients from only Spain, for instance, showed such a phenotype.

The challenge is that many PRDs lack appropriate registriesdsometimes, registries are established, for instance, by academics or by patient organizations, when a pot of money becomes available via a grant, for instance. Too often in this scenario, registries are not sustainable and fail to be curated and updated. In December 2021, Europe alone had over 800 separate registries for rare disease (Taruscio et al., 2013). Many of these, even if they are still maintained and updated, were set-up prior to the emergence of data science and, in particular, FAIR data principles. These early registries likely have limited resources to *become* more interoperable, which limits the value of the legacy data. For *very* rare conditions, the natural history of which is very poorly understood; it may be that mining health records, by making them more interoperable, for instance, by promotion of a European EHR exchange standard, would yield a more accurate picture than structured, limited, registry datasets. This is because for a registry to be a reliable source of natural history data in a rare disease, one needs to have an idea of the likely symptoms and organ systems involved, in order to incorporate them to the dataset to monitor them. Arguably, therefore, it is more difficult to detect unknown impacts of a condition in this way.

Again, many recommendations have been proposed, and many projects are underway, designed to increase the power of registries, electronic health records, and other sources of data, to elucidate natural history and accelerate research into hitherto neglected conditions. The following are a very few examples from the Rare 2030 recommendations, to illustrate the kinds of steps and strategies proposed to European and national actors.

- A renewed multistakeholder dialog is required, at the regional, national, European, and global levels, to ensure a more strategic approach to the creation and connectivity of rare diseases registries and data repositories, at all levels.
- European reference networks should be specifically and adequately funded to develop and conduct natural history (and where possible accompanying biomarker) studies, a minimum of 5 every 2 years, to build the knowledge base and capacity for clinical research in disregarded diseases/areas lacking research

Once a community has reached these points, that is, there is sufficient understanding of the condition, a number of additional challenges present themselves, around the concrete planning and delivery of a trial.

## Challenges in designing and delivering PRD clinical trials and current solutions to address these

Fig. 19.2 summarizes the major challenges in the design and delivery of PRD trials along with potential solutions. These challenges are discussed in detail below.

### The rarity of disease and corresponding scarcity of patients makes planning a trial and finding expertise difficultdwhat solutions exist today?

Just as the rarity of the disease can hamper *basic* research for many rare diseases, that same rarity poses challenges for planning and initiating a clinical trial, challenges which do not tend to appear in common conditions. Sponsors planning a clinical trial need to know which sites will be able to attract a satisfactory number of participants, no easy feat if only a handful of centers across Europe regularly treat patients with that condition. Fortunately, the PRD community now has a number of infrastructures and initiatives which are improving the findability of sites and patients. To make further progress, it will be necessary to consider how some of these practically and tangible interact to make it as easy as possible to find and recruit patients.

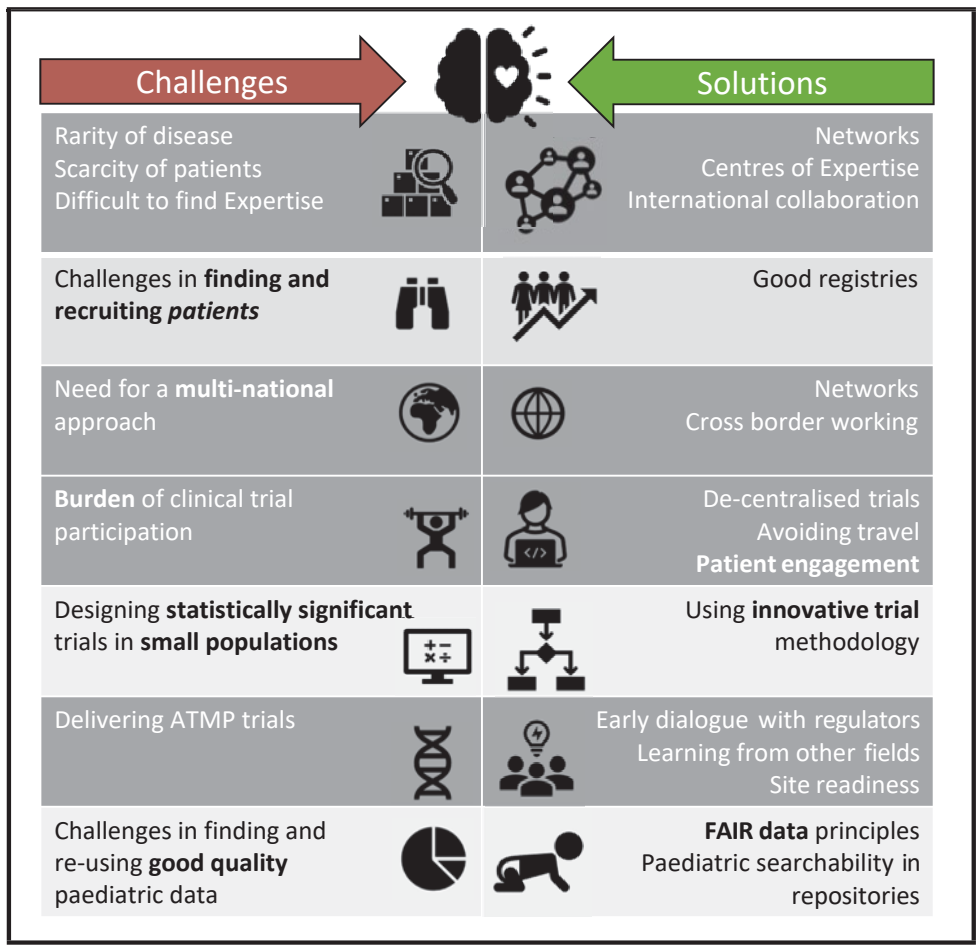

FIGURE 19.2 Challenges in designing and delivering pediatric rare disease clinical trials and potential solutions.

*Centers of expertise*

Back in 2011, the expert group for RD appointed at European Level (EUCERD) unanimously adopted a set of recommendations on quality criteria for centers of expertise for rare disease. These were ambitious but carefully selected criteria intended to help countries map what RD expertise existed in their country and where. This was intended to help EU MS fulfill the request issued to them by the 2009 Council Recommendation on an action in the field of rare diseases: "*[MS should] Identify appropriate centers of expertise throughout their national territory by the end of 2013 and consider supporting their creation.*"

The general expectation was that countries would use those EUCERD recommendations as their baseline criteria, perhaps supplementing them with disease-related criteria to distinguish the different sort of expertise a metabolic center would need compared with a neuromuscular center or a center of expertise claiming to specialize in intellectual disability syndromes. This was a real opportunity to give visibility to the national picture, helping multiple stakeholders (whether patients seeking a diagnosis or general clinicians seeking an expert center or a sponsor seeking experts and sites to participate in a clinical trial) to overcome some of the challenges of rarity.

*European Reference Networks*

The emergence of ERNs in 2017 represented a major watershed moment for rare diseases in Europe generally but also constituted an important turning point for finding experts and patients more easily. ERNs were born from the principle that no single member state alone can solve all the issues related to rare disease diagnosis, treatment, and care.

In February 2017, 24 ERNs were launched, embedded in a European legal framework (the so-called cross-border Healthcare Directive 2011/24/EU). They were established under thematic groupings, at a high level, designed to provide a "home" for every rare disease. As of January 2022, 1450 HCPs (typically highly specialized centers or units within larger hospitals) participate as full members of the 24 ERNs. Although primarily healthcare focused (in particular perhaps, providing highly specialized care across borders), and making strides in capacity-building and education, the ERNs have a requirement to facilitate -indeed conduct-research. This is a logical role and echoes the dual responsibilities of any so-called expert center: the care and research boundary is quite blurred in rare disease, and true expert centers need to be able to engage in research and knowledge generation as well as care provision. The ERICA project (Desvignes-Gleizes et al., 2022, p. 25) is dedicated to advancing ERN research, in all its forms. Together4RD (https://together4rd.eu/) is seeking to advance ERN-Industry collaboration, in particular, which has been limited to-date.

Although the ERNs are appointed with broad remits, for example "rare endocrine," they established numerous clinically relevant subdomains. As part of the original applications to become an ERN, each had to stipulate disease-related criteria for any HCPs wishing to join. These "vertical" criteria in fact constitute a robust pan-European attempt to define what truly constitutes expertise in, for instance, inherited metabolic diseases or rare eye disease, etc. These criteria specify patient numbers, necessary equipment/procedural skills, and the medical, multidisciplinary, and paramedical expertise that should somehow be accessible by any HCP claiming expertise in each domain. As such, they hold major potential for countries perhaps seeking to determine disease-related criteria to supplement the cross-cutting EUCERD criteria of 2011 on Centers of Expertise for RD.

Where ERNs have made concerted efforts to map (European Reference Networks) the expertise offered by their constituent national HCPs, this information can be made available and could be a very valuable resource to support trial sponsors in finding expert centers likely to possess both specialized knowledge to run trials but also to see a reasonable proportion of the patients with a given condition that falls under their area of declared specialty.

*conect 4children*

conect4children (c4c) is a public-private partnership funded through the innovative medicines initiative to establish a pan-European clinical trial network to address barriers and provide an infrastructure for to deliver better, more effective clinical trials for the pediatric population (Turner et al., 2021). The ultimate aim of c4c is to create better medicines for babies, children, and young people. The c4c network supports sites to setup and deliver trials through coordination by NH. National hubs meet the specific needs of sponsors, contract research organizations, and academic investigators during trial setup and conduct. Support includes standards for trial conduct, metrics for trial performance, and a quality framework. NHs bring benefits as they can act as a national point of contact for c4c; supporting sponsors in coordinating trial conduct in their country, they may support sites in discussions about contracts, regulatory, and ethics approvals by aligning templates and providing mediation when needed; managing c4c feasibility requests at country level and supporting sites during clinical trials (Turner et al., 2022). One advantage of the NH is that over time, these hubs should build knowledge about who is working in which pediatric diseases at different sites across any given country.

c4c has brought a greater level of organization to the European pediatric research community, specifically, which is also helping to provide a clearer picture of who is working in which rare pediatric disease, and to what level of ability. Patient involvement is central to the vision of c4c, including the views of children, young people, and their parents in the design and conduct of clinical trials is critical to improving the delivery of pediatric clinical research. Patients and their families are also involved in providing advice to sponsors to co-design studies for children. This allows sponsors to optimize study design/pediatric development plans through expert advice from clinical and methodology experts as well as patient/carer representatives, for both industry and academia (Sen, Palmeri, et al., 2023).

*Pooling the knowledge and resources of the rare and pediatric fields*

The focus and scope of the aforementioned activities and tools developed by ERNs and c4c are not identical. However, there are important commonalities. ERN member centers will, in some cases, be more care-focused than research-focused (this is particularly the case for ERNs which represent historically less research-active communities, which have not secured major research and networking grants in the pre-ERN era, etc). But given the research criterion, these will often house key expertise essential for clinical trials, and will have access to patient cohorts (see below). Now, in many countries, clinical trials in pediatric populations will need to involve centers beyond those which are officially members of any given ERN—but often, pediatric clinical trials may end up involving ERN HCPs. Furthermore, ERNs are explicitly supposed to address both pediatric and adult conditions under their Thematic Grouping, which means that PRDs should be "on the radar" of both the ERNs and the trial landscape being developed under c4c. In future, as the NHs become more active in supporting assessments of the feasibility of clinical trial sites within the national territory, they should increasingly look to the mappings performed and shared by ERNs, to ensure that for *rare* pediatric conditions, they know where the top expert sites are located. Combining the work performed by c4c with that of ERNs should make it easier to locate real experts working in different PRDs.

Challenges in finding and recruiting patients—what solutions exist today?

In rare and pediatric diseases, clinical trials often fail due to an inability to recruit the requisite number of patients. This is unsurprising as the rarer the condition, the smaller the potential patient pool to participate in a clinical trial. An increased ability to more easily find centers of expertise specializing in PRDs, as outlined above, will logically support the finding of patients. This is because *true* expert centers, spanning the clinic and the research arenas (as called for in the EUCERD Recommendations, and as incorporated to the Delegated and Implementing Acts upon which ERNs are based) naturally become a conduit for people diagnosed with—or suspected of having—a condition falling under that sphere of expertise, for example, a rare bone disease, broadly, or a particular condition like Osteogenesis Imperfecta. They naturally maintain records of patients, which they can use to contact prospective trial participants when approached by a pharmaceutical company to participate in a clinical trial.

Another useful tool when it comes to finding patients for clinical trials in PRD is a robust, up-to-date registry. A center of expertise (such as an ERN HCP) tends to become the key referral point for patients with a given PRD, and as such its experts are often aware of the existence of suitable registries, in which their patients may be enrolled. This may be a registry their center actually runs, or else—more commonly—staff at such a center will routinely offer patients the chance to register in a suitable national, international or global registry. There is a great variety in the way that registries are structured and managed: they can be patient-reported, clinician-reported, or a combination of the two. Some are run by patient organizations, and some are integrated into the health care system. The diversity in national registries can pose a challenge, but creating a shared dataset and data dictionary across registries can support interoperability (dos Santos Vieira et al., 2022; Sen, Hedley, et al., 2023). The development of shared datasets should be done in collaborative ways to ensure a multistakeholder perspective, including the patient voice.

Finding patients is therefore another potential role for a robust registry—depending on how a registry is established (i.e., on the data it collects, and how frequently), it can support an understanding of natural history, generate knowledge on "what works" clinically (e.g., correlating outcomes such as ambulation with use of corticosteroids in Duchenne Muscular Dystrophy), and support feasibility assessments for clinical trials. A good example of the latter comes from the neuromuscular community. Established in 2007, TREAT-NMD is a network of excellence for rare, inherited NMDs. TREAT-NMD has played a central role in bringing together the right experts, patients, advocacy organizations, scientists, healthcare professionals, and pharmaceutical companies, to advance trial readiness (Leary et al., 2017). For instance, TREAT-NMD has successfully developed several tools and key resources, including cell and animal standard operating protocols (preclinical research), global patient registries, care guidelines, and family guides, to help develop and extend translation research in the field, thereby making the field ready for clinical trials. Investment in disease specific, national registries, with a harmonized data set, allows for a global patient cohort, where data can be accessed and analyzed to support research about the disease, trial feasibility, and patient recruitment (Bladen et al., 2013). In terms of increasing the findability of patients for clinical trials, industry has long-since used the tools and procedures developed by the TREAT-NMD network to distribute queries to independent but federated registries in the neuromuscular field, seeking aggregate data on, for instance, the number of patients with condition X living within 100 miles of a given set of trial sites, who fulfill certain inclusion criteria. Through a system of checks and balances (in which the request is reviewed by a TREAT-NMD Global Data systems Oversight Committee (TGDOC) composed of curators of all the standalone registries), the global registries for conditions like DMD and spinal muscular atrophy can advance clinical trials by de-risking planning for Pharma. They can also facilitate the actual recruitment of patients, should a study go ahead, again without any direct contact between pharma company and individual patients and families.

In fact, the neuromuscular community under TREAT-NMD developed a further tool to support the finding of patients and increase the feasibility of planning a clinical trial—the TREAT-NMD Care and Trial Site Registry (CTSR) maintains a record of sites, which include the number of patients with each neuromuscular disease who are attached to given centers/hospitals across Europe (Rodger et al., 2013).

While many PRDs still lack clinical research opportunities, some conditions have seen significant increases in clinical trial interest, which actually bring *new* challenges of their own. Some communities are innovating in an attempt to address some of the challenges around finding patients in conditions with a thriving R&D ecosystem.

*DMD Hub Case study—expanding capacity and expertise for trials in rare disease*

A detailed description of a DMD Hub case study is presented in Box 19.1.

---

BOX 19.1

### DMD hub case study

Duchenne muscular dystrophy is one of the more common rare diseases affecting around 1 in 3500 male births worldwide. Following the discovery of the dystrophin gene in 1987, there were no treatment options for the disease, other than corticosteroids, and a distinct lack of clinical trials offering hope. However, from 2005 onward, interest from the pharmaceutical industry, and subsequently the number of clinical trials in DMD, have steadily increased.

In 2015, the two leading UK neuromuscular centers of excellence (John Walton Muscular Dystrophy Research Centre, Newcastle University and Great Ormond Street Hospital, University College London) were almost exclusively the only UK sites running clinical trials in DMD and were reaching capacity. It became clear that this was impacting on the ability of the increasing number of trials wanting to come to the UK, being able to be set up.

As a result, the DMD Hub was established as an innovative collaboration between Duchenne UK and leading UK neuromuscular centers of excellence. With investment exceeding £3.5 million from Duchenne UK, 34 posts have been funded at 11 sites resulting in the establishment of a trial ready network. The DMD Hub has succeeded in increasing capacity across the UK and has enabled more sites to take on additional trials and to recruit over 400 boys to DMD clinical research since 2015, providing additional opportunities for patients to participate in research.

However, DMD research remains at an unprecedented stage in terms of the number of potential therapies coming to trials (with at least 32 trials running or in the pipeline). With recent advances including the arrival of gene therapy and the expansion of trials into the older nonambulatory population (rather than exclusively in the pediatric population), the DMD Hub has expanded its focus to prepare for gene therapy trials and include adult sites. These are both relatively new areas to the neuromuscular field and the DMD Hub ensures that there are sites with capacity and expertise ready to deliver the new generation of clinical trials, providing even more opportunities for DMD patients to participate in research.

Identifying issues, sharing expertise, and providing training and education were all fundamental to developing and further expanding the DMD Hub network. The DMD Hub has developed staff support networks for nurses and clinical trial coordinators and resources to facilitate working together in order to upskill sites and share expertise. Key examples of these sources of support are the Toolbox and Training platform, which are hosted on the DMD Hub website (www.dmdhub.org) as well as coordination of staff network meetings and annual training courses.

> BOX 19.1 (cont'd)
>
> ### DMD hub case study
>
> Collaboration between clinical trial sites, industry, and patient organizations, as well as other relevant national and international initiatives such as the National Institute for Health and Care Research (NIHR), Northern Alliance Advanced Therapies Treatment Centre (NA-ATTC), and TREAT-NMD has been key to the success of the DMD Hub. In the field of rare diseases, it is essential to ensure that all stakeholders are working together, not duplicating efforts or expending precious resources, particularly patients willing to participate in trials.
>
> Interestingly, however, an increasing number of clinical trials opening in the UK, providing more opportunities for patients to participate in research at a still limited number of sites, brings with it other dilemmas:
>
> For patients/parents, there are difficult decisions to be made when deciding which trial they may be interested in participating to. Patients and parents also want reassurance that there is fair and equitable access to trials across the UK, and not living very close to top centers does not pose a disadvantage for families who are often desperate to access treatments in which so much hope is placed. However, in attempting to open up access to trials more fairly, it is important to ensure we do not limit participation to those patients receiving their usual care at the sites running the trials.
>
> The growing clinical trial scene for DMD also poses challenges for sites and industry, who need to ensure that the required number of patients are available to participate in the trials, in order to meet recruitment times and targets.
>
> To inform patients of trial opportunities in the UK as well as support equitable access to clinical trials, the DMD Hub developed the clinical trial finder (CTF). The CTF lists all current and upcoming trials along with the recruitment status at the sites running the trials so parents and patients can be aware of opportunities. In an additional effort to address timely and effective recruitment and match interested patients with trial places available, the central recruitment pilot study was launched and has developed a national procedure for recruiting out of area patients to clinical trials (https://www.dmdhubrecruits.org/). This model is currently being tested and initial indications look positive.
>
> Managing expectations of parents and patients of the potential therapeutic value of gene therapy is also something the DMD Hub in collaboration with Duchenne UK is working to address. Providing clear and consistent communication from all stakeholders (including patient organizations, industry, and clinicians) regarding the potential of gene therapy and the rationale for clinical trials is important. Webinars, newsletters, and patient information days are particularly useful ways to disseminate the information. As is ensuring that staff working on the trials are adequately trained and supported to deal with the potentially difficult conversations and explain complex concepts to families.
>
> Working collaboratively across the DMD Hub network facilitates sharing and increases the knowledge within the DMD community, helping to ensure we are in as strong a position as possible to effectively and efficiently deliver clinical trials.
>
> The results, methods, and resources developed within the DMD Hub are being shared so that other countries and other disease areas may also benefit.
>
> In the DMD research community, every patient counts and increasing access to trials and ultimately treatments is vital.

## Challenges in setting-up and delivering multinational trials in PRDsdwhat solutions exist today?

Multinational clinical trials are essential in rare diseases research. They provide greater access to patients, facilities, and medical expertise and can overcome the scarcity of patients. A mentioned above, the 10-year report on the European Paediatric Regulation identified several major challenges in the conduct of pediatric clinical trials (European Medicines Agency and its Paediatric Committee, 2016). While some of these are generic for all pediatric trials, some are specific for multinational trials. These include differing regulatory processes and timelines, site-specific contracts, different ages of consent (or parental consent), different ethical standards, and differing institutional review boards (Caldwell et al., 2004; Lim & Cranswick, 2003). These challenges massively slow down the initiation of a trial. For example, palatability of drugs for children is an important ethical issue. However, there is no gold standard regarding this (Batchelor et al., 2015). Hence, subjective judgments are made, and invariably, there is a possibility of the ethical process hitting a roadblock in one or more countries. The aforementioned challenges often affect pediatric rare disease disproportionately due to even smaller populations. For example, a recent study indeed highlighted the wide range of informed consent ages prevalent across Europe (Lepola et al., 2016). Based on this, if the age limit of a multinational trial is set to 12 and over to accommodate age of consent laws in multiple countries, this will exclude younger children from the countries where the age of consent is lower. c4c has developed multiple tools and resources to facilitate and expedite the start-up of trials in pediatric diseases, including the following.

### Expert advice

The c4c Expert Network exists of more than 300 experts organized into clinical subspecialty groups, innovative methodology groups, and patient/parent representatives. During the course of pediatric drug development, investigators, either from industry or academia, can contact c4c to ask for strategic feasibility advice. This advice can encompass clinical questions, methodology, and/or patient and public experts and is provided through live or online meetings. In these expert meetings, involvement of patients and public reaches much further than advice on informed consent and assent forms and feedback of study results but encompasses input in the actual study design including medical need, study endpoints, study procedures, and planning of visits (de Wildt & Wong, 2022).

### c4c academy

The academy offers a range of online courses to educate and train study trial teams to deliver more effective and efficient clinical trials. This includes courses on basic and advanced aspects of GCP, PPI, study conduct, and scientific short courses, all of which focus on the need for and challenges of multicenter pediatric clinical trials.

### Patient and public involvement theme

Involving patients and their families in the design and delivery of pediatric clinical trials is a core value of c4c. The pan-European nature of c4c ensures that the views of parents and their families from different countries and cultures are included. Research should be designed in such a way that the needs of those taking part in research are met. PPI is a fundamental element of good practice in research (Preston et al., 2019).

### ECRIN

ECRIN was originally established under the ESFRI roadmap. ECRIN is now a European research infrastructure that facilitates researchers to set up and conduct multinational clinical trials in Europe. This is achieved through linking national networks of clinical trial units and through the provision of services and tools (Demotes-Mainard & Kubiak, 2011).

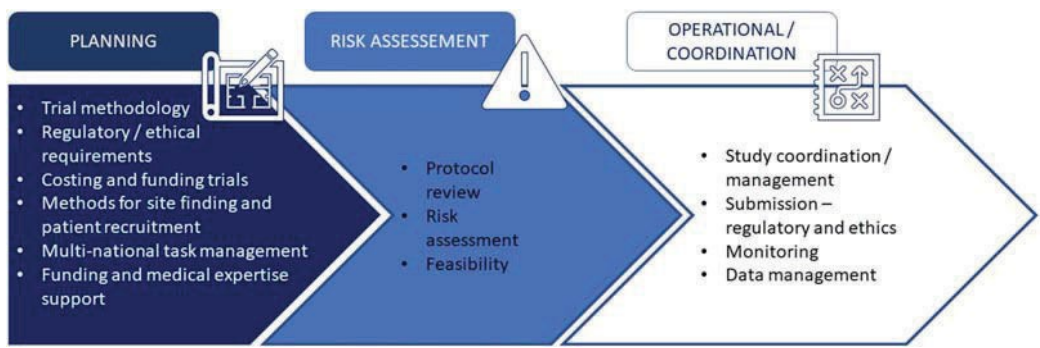

FIGURE 19.3 Some examples of clinical trial applications where ECRIN toolkits could be used. *The figure is adapted from a similar figure on the ECRIN website (https://ecrin.org/clinical-trials).*

ECRIN focuses its work on the conduct of investigator-initiated trials, across all disease and topic areas. This includes trial set-up and management in the multinational setting; this is especially important for rare and pediatric diseases. Services are offered from the trial planning and design phase, through to operational trial management. ECRIN provides a toolkit to support sponsors with the various stages of planning and delivery of multinational trials and provides examples of clinical trial applications where ECRIN toolkits could be used (Fig. 19.3).

## The need to make the experience of participating in a trial as easy as possible for PRD patients—what solutions exist today?

To make participation in a clinical trial as easy as possible for young patients and their families, it is important to ensure that the views of parents, carers, and children are taken into account when designing a trial. Patients' and parents' contributions to the development of pediatric studies are also critical to ensure that end points and assessments are meaningful and feasible.

Networks such as c4c can have a key role here through providing advice and support where views of children, young people, and their families are central. c4c can promote innovative trial design to optimize pediatric development plans and protocols through the provision of expert advice to sponsors. The goal of advice is to meet the needs of the sponsor and children. To ensure the views of patients and their families are integrated, c4c has created a database of patients (Cheng et al., 2023). This database contains 135 registrants, as of September 2022. Four subgroups are defined within the database: *"(1) professional of a patient organization, (2) YPAG facilitator, (3) adult patient, and (4) carer of a young patient."* Patients under 18 years of age are not permitted to register themselves.

Although the COVID-19 pandemic brought multiple challenges to the delivery of research and healthcare, it also offered opportunities to introduce innovative practice and increase the number of DCT. The FDA and EMA definitions of DCTs are provided in Table 19.1.

Greenberg et al. provided a list of clinical trials elements that can be conducted remotely or away from the traditional clinical trial setting. They include.

- online recruitment
- home visits to dispense the IMP or take samples for laboratory tests
- direct shipment of study materials to the participant's home, including options for electronic consent/assent

TABLE 19.1 Definitions of decentralized (remote) clinical trials (taken verbatim from EMA/FDA websites).

| EMA | FDA |
| --- | --- |
| The aim of DCTs is to make it easier for patients to participate in clinical trials by reducing the need to travel to central trial sites. Decentralization is enabled by the advancement of digital tools, telemedicine and more mobile and local healthcare. It includes aspects such as home health visits, remote monitoring and diagnostics, direct-to-patient shipment of study drugs and electronic informed consent. (https://www.ema.europa.eu/en/news/facilitating-decentralised-clinical-trials-eu) | The remote collection of trial data outside of a standard in-person clinical trial site. These methods include patient-centric trial flexibilities such as electronic informed consent, virtual clinic visits, delivery of investigational product to the home and obtaining laboratory or imaging assessments locally.(https://www.fda.gov/about-fda/oncology-center-excellence/advancing-oncology-decentralized-trials) |

- evaluations by video including assessment of outcome measures.
- telephone calls for clinician appointments and safety screening
- online assessments which can be used for patient reported outcome measures
- use of wearable devices and mobile apps to collect data

By offering fully decentralized trials, or elements of decentralization (hybrid) within clinical trials, it can reduce the burden on the family of taking part in clinical trials. The decentralized approach can result in a reduction in the number of visits to hospital. This can bring many benefits to the child and family including a reduction in days missed at school, reduced impact on family life, and a lower caregiver burden. All of which can improve recruitment and retention within a clinical trial.

The use of DCT is likely to increase over the coming years. The EMA has recently released a recommendations paper on DCT (Directorate-General For Food and Health Safety, 2022), and the FDA released guidance during the pandemic (FDA, 2020). DCT offer great opportunities to improve recruitment for pediatric rare disease trials.

## The need to design statistically significant trials in small target populations—what solutions exist today?

While attaining the required level of statistical power with the minimum number of patients is important for *most* trials, this is critical for pediatric and rare disease trials. Given the scarcity of rare disease and pediatric patients, and the challenges in recruitment, the use of emerging new trial designs holds significant potential to improve the planning and delivery of such trials. Designs of particular interest (many of which stem from the oncology field) include the following.

### Multiarm designs

Multiarm trials benefit from the use of a shared control group, resulting in the equivalent statistical power being generated from a smaller number of patients. For example, two separate two-arm clinical trials recruiting 20 patients in each arm would require a total of 80 patients. A three-arm design with a shared control group would generate the same statistical power with 60 patients (in practice, a slightly higher sample is required after accounting for multiple-testing corrections (Odutayo et al., 2020)). This kind of design is popular with patients (or parents) who want access to the experimental treatment, as they are less likely to be in the control group than on a traditional trial (Jaki & Wason, 2018). For sponsors and administrators, multiarm trials are easier to run than multiple two-arm trials. An additional advantage is the possibility for a "head-to-head" comparison between two treatments. One downside, however, is that patients (or parents) may be more inclined to decline to take part if they are uncomfortable with any one of the treatments. For this kind of trial design to work, each treatment will need to be available at each site. Examples of multiarm designs are NCT00733096 (Mulleman et al., 2006) (3-arm) and NCT01076699 (Hardi, 2010) (4-arm).

*Platform designs*

A platform design is similar to a multiarm design insofar as it also uses shared control arms, but in this methodology, arms can be terminated or added to that starting structure, following interim analyses. A shared control arm in a platform trial offers similar advantages to those in multiarm trial designs outlined above. For example, a five-intervention platform trial requires 45% fewer patients than five two-arm trials (Gold et al., 2022). While the interim analysis must be of a high standard, platform trials offer the opportunity to stop ineffective treatments early. The addition of new arms can enable a treatment to reach patients more quickly, without having to repeat the lengthy ethical, regulatory, and other approvals necessary for a whole new trial. One consideration for adding new arms is whether to use concurrent or nonconcurrent controls. While nonconcurrent controls provide a greater chance to detect differences, there are likely to be period effects (changes over time). Examples of platform trials are the I-SPY 2 trial (Esserman et al., 2012) and the STAMPEDE trial (James et al., 2009, 2017). These trials have been ongoing for over a decade and have tested several treatments for breast and prostate cancer, respectively. Arms are continuously added and removed. The benefit of platform designs has been highlighted by EU PEARL—a public-private partnership to shape the future of clinical trials—which has performed case studies in COVID-19 and nonalcoholic steatohepatitis (Pericàs, Derde, & Berry, 2023; Pericàs, Tacke, et al., 2023; Roig et al., 2022).

*Factorial designs*

A factorial design can potentially combine two or more different trials into a single trial, thus substantially reducing the number of patients required to attain statistical power in both. This design is particularly useful if the two treatments are known not to interact (either positively or negatively) (Montgomery et al., 2003). A key consideration, however, is that the prevalence of interaction must be known beforehand. If an interaction is expected, the sample size needs to be increased appropriately to maintain statistical power (unless the interaction effect is expected to be at least twice as large as the main effect) (Freidlin & Korn, 2017). An example of such a trial is the PRESERVE trial (Garcia et al., 2018) that tested two treatments for the prevalence of serious adverse events following angiography. The treatments could be tested separately as two two-arm trials—saline versus bicarbonate and placebo versus n-acetylcysteine. However, under the factorial design four arms were created—saline + placebo, saline + n-acetylcysteine, bicarbonate + placebo and bicarbonate + n-acetylcysteine. Despite the four arms, the primary comparison remained the same as the two two-arm trials. If the main interest is in fact the combination of the two treatments, a multiarm trial may be more suitable.

*Crossover designs*

Unlike two-arm or multiarm designs where each arm receives one type of treatment (or placebo, in a crossover design), each patient is administered all the treatments successively. Hence, each patient acts as their own control. For a two-arm trial, the number of patients required to maintain statistical power is halved. In fact, a recent review found the median sample size of crossover trials to be 15 (Mills et al., 2009). In addition, the between-subjects variability, which is highly prevalent in two-arm and multiarm designs, is essentially eliminated. Disadvantages of this methodology include the requirement for patients to be at the same level of health when commencing each treatment. This necessitates a washout after the completion of each treatment which ensures that there are no carryover effects from the previous treatment. This requirement also reduces the scope of such trials to target chronic disease. Finally, the successive nature of treatment administration can lead to period effects (e.g., seasonal effects). Examples of a crossover studies are NCT02518230 (Rosterman et al., 2018) and NCT03580707 (Reed et al., 2020).

*Basket and umbrella designs*

Basket and umbrella trials can investigate multiple diseases (or multiple forms of the same disease) simultaneously. Diseases with shared underlying genetic mutations or potentially similar mechanisms of action are studied together in basket trials, including when these clusters of diseases have different clinical manifestations. The same treatment is provided for all diseases (McNair, 2020). The Pembrolizumab trial is an example of basket design, which included locally advanced or metastatic carcinoma, melanoma, or non-small-cell lung carcinoma (Robert et al., 2014).

Umbrella trials include patients with different genetic mutations for the same disease. The treatments are usually mutation specific. An example of an umbrella trial is plasmaMATCH where five different therapies were provided to 5 treatment groups based on their molecular signatures (Turner et al., 2020). Both these types of trial designs have mostly been used for multiple types of cancers.

The causes of rare metabolic disorders often lie within the same biomolecular pathways, making them strong candidates for basket and umbrella trial designs (Liu et al., 2021). Basket trials bring many advantages; for instance, they can reduce the costs associated with trial set up and recruitment. Their design means fewer patients with each individual disease need to be recruited, alleviating traditional recruitment challenges. The utility of basket and umbrella designs for rare disease are discussed by Newton et al. (Newton, 2021).

The basket methodology has a further use case in pediatric clinical trials: there may be good amounts of safety and efficacy data about a therapy from adult studies, so if this same drug is then tested in children, data from the adult study can be borrowed/reused to inform pediatric studies.

## Challenges in designing and delivering clinical trials of ATMPsdwhat solutions exist?

Gene therapy modifies a person's genes to treat, or ideally cure, disease. It can do this in a number of ways depending on the type of gene therapy and the therapeutic indication. Alongside somatic cell therapy and tissue engineered products, gene therapy is classed as an advanced therapy medicinal product, or ATMP. As around 75% of rare diseases are believed to be caused by a genetic defect, gene therapy in particular shows great promise for treating many rare pediatric conditions. Research in this sphere is accelerating at pace, with several gene therapies now becoming licensed medicines and the number is expected to continue to rised200 ATMPs are currently in phase three trials, with approximately 2000 in development. Such therapies hold the promise of effectively treating conditions which, otherwise, would prove fatal in the first years of life, such as SMA and metachromatic leukodystrophy (MLD).

There are particular challenges, however, in planning and delivering clinical trials in such therapies beyond what is noted in an "ordinary" Clinical Trial of an Investigational Medicinal Product (CTIMP). For instance, ATMPs will typically incur a need to navigate extra approvals and notifications. Site-specific risk assessments will need to be performed, to put in place the requisite measures for handling and delivering gene therapy products, and additional/specialist training for clinical trial staff and waste management may be needed. It may be trickier to coordinate a complex visit schedule with intensive safety monitoring and a lengthy follow-up period. The set-up and delivery of these trials can therefore be time consuming and involve coordination between many different individuals. These trials also need to be appropriately costed, taking into account the additional considerations involved. These challenges, on top of the challenges of performing a CTIMP in a PRD generally, can be very significant.

The additional complexities involved in ATMP trials therefore necessitate careful prior planning to ensure the appropriate infrastructure, resources, and capacity are in place: a thorough institutional readiness assessment, alongside a detailed feasibility review, is required at an early stage, to address the enhanced requirements of these complex trials. Hospitals involved in ATMP trials will have a genetic modification safety committee (GMSC) in place to assess the risk associated with conducting an ATMP trial. Clinical investigators and trial sponsors need to provide relevant information for the GMSC to make an informed decision about the feasibility of ATMP trials.

The first communities to trial and approve gene therapies have valuable experiences to share, in terms of practical advice and pitfalls to avoid (Heslop et al., 2021; Stirnadel-Farrant et al., 2018). For instance, in the Duchenne muscular dystrophy community, the DMD Hub is a good example of a (UK) charity-funded initiative, which is pioneering in the gene therapy field when it comes to rare disease research. The DMD Hub provides a national network for sites running trials in DMD and disseminates guidance and resources specifically for gene therapy. There may also be national bodies and initiatives offering expertise in this relatively new area of research. For example, in the UK, the National Institute for Health and Care Research (NIHR) provides vital infrastructure and support for delivering research generally, and the Cell and Gene Therapy Catapult offers a wealth of resources to support the delivery of ATMP trials, specifically.

Naturally, advice around development of ATMPs and GCP specificities in Europe is provided by the EMA. But to address some of the challenges observed by the first line of communities engaging with ATMP trials, it makes sense to leverage experience, which exists from individual communities, imparting valuable learning lessons to other disease areas: there is a real benefit to cross-disease initiatives or expert bodies able to impart such advice *across* the PRD domain. For instance, the Expert Groups established by c4c may be able to advise on practical issues of relevance for pediatric trials in ATMPs, as part of their consultancy service (see 18.3, above).

Cross-disease initiatives more focused on the downstream topic of *access* are also relevant here. This is very important, as although arguably *any* clinical research in PRD should be considered as part of a cohesive research ecosystem (with consideration given very early in the therapy development process to market access and postmarketing surveillance needs), it is especially important to factor-in the downstream challenges of bringing an ATMP to patients with a rare disease. The uncertainties around long-term use of ATMPs, the high cost of these therapies, the fact that benefits may be less immediate from some ATMPs than is traditionally desirable—all are likely to pose particular concerns for HTA bodies and payers (Horgan et al., 2020; Morrow et al., 2017). The Rare Impact project explored and summarized challenges and solutions around access to gene and cell therapies across Europe (EURODIS, 2020). More recently, in 2022, an initiative called AGORA was launched, engaging clinical academics, scientists and patient organizations, with the goal of providing support and advice for not-for-profit organizations seeking regulatory approval for ATMPs intended for rare disease, which have proven to be effective but are not commercially sustainable.

## Challenges in finding and reusing good quality pediatric data—what solutions exist?

As pediatric clinical data are scarce, making use of existing data from clinical trials, registries, observational studies, and electronic health records can be extremely useful for planning and delivery of trials in PRD. Some potential reuses of data are discussed below.

### Patient recruitment

Projects such as the IMI-funded EHR4CR project succeeded in combining and extending several previously isolated components to develop a platform for reusing EHR data to support medical research (De Moor et al., 2015). This platform included support for patient recruitment through EHR (Dupont et al., 2018).

### Use in preauthorization

Use of data in preauthorization is less understood, but a recent review of 111 medicinal products authorized by the EMA identified the following uses (Eskola et al., 2022).

- Identification of patient populations
- Understand disease features
- Assessment of the burden of disease
- Informing trial design (less common)
- Supporting assessment of efficacy/safety
- Support comparisons to the current clinical practice in early medicines development
- Compare therapeutic benefit and effectiveness between new and currently available treatments

### Innovative trial design

Using relevant individual patient data on control from external trials, or real-world data sources, may allow the reduction or even elimination of the concurrent control group (Schmidli et al., 2020). This could be of particular importance to pediatric rare diseases where the populations are small and many diseases are progressive and or fatal in childhood.

### Postmarketing surveillance

Reused data can be utilized for postmarketing surveillance. In Europe, the ENCePP aims on conducting postmarketing risk assessment using various EHR sources (Goedecke & Arlett, 2014, pp. 403e408). In the United States, the FDA is responsible for postmarketing surveillance and uses EHR data from several different sources (Cowie et al., 2017; Staffa & Dal Pan, 2012).

### Pragmatic clinical trials

Electronic health record systems can now support RCT. In some cases, an RCT may be embedded in the EHR and healthcare delivery system. This can be done so effectively, that is, without interfere with clinical practice in any way and can remain almost "invisible" to the clinician (Pletcher et al., 2020).

Pragmatic clinical trials are those that aim to show the effectiveness of treatment options in a real-world setting. EHRs are extremely useful to the effective delivery of pragmatic clinical trials (Simon et al., 2018).

### Comparator arms

A potential way to reuse pediatric data is to use RWD as a comparator arm for clinical studies. C4c is currently exploring the feasibility of using data found in EHRs to support clinical trials, as well as trying to identify what the barriers may be. A barrier for comparator arms is that RWD in form of EHRs use a different format to clinical studies. Clinical studies are subject to regulatory requirements, and CDISC standards are mandated in USA by the FDA and Japan by the PMDA (Bobbitt et al., 2019; FDA, 2017). CDSIC standards are also recommended by the EMA in Europe and by NMPA in China. The most common formats for EHRs are FHIR and OMOP. There have recently been multiple efforts to translate CDISC standards to OMOP/FHIR and vice versa (CDISC, 2021; Davydov et al., 2020; Facile et al., 2022). However, as with most translations, a loss of information is associated with the process.

*Repositories*

As mentioned above, the FAIR principles are crucial in leveraging the maximum value out of data. This includes provisions to use quality data from past studies. Due to the general scarcity of pediatric data, attaining past data could fill important gaps in current research. Several repositories storing clinical trial data have been made publicly available—YODA, Vivli, RDCA-DAP, etc (Felisi et al., 2024). However, there remain challenges in using these repositories for research. While all these repositories accept pediatric studies, the number of pediatric studies is generally quite low. For example, less than 10% of YODA's 449 studies (as of February 17, 2023) have a median age below 19. Moreover, despite being publicly available, the data request and approval process often runs into several months and are generally not useable for short-term grant-based projects. Lastly, though all the repositories are committed to implementing the FAIR principles, the use of legacy data greatly escalates the cost of FAIRification. As a result, substantial portions of the data are unformatted and lack interoperability.

*Working collaboratively to solve the challenges*

Within c4c it was recognized that the challenges to find and harmonize rare and pediatric data cannot be solved by just one project. Therefore, it was agreed to forge a collaborative forum across multiple project working in the health data space. The GLOPAD (Global Pediatric Data) Forum was first convened in May 2022 (Sen et al., 2024).

Key objectives of the group (Fig. 19.4).

- Avoid duplication in the field.
- Ensure c4c can address the pediatric-specific aspects of data harmonization, by partnering with others.

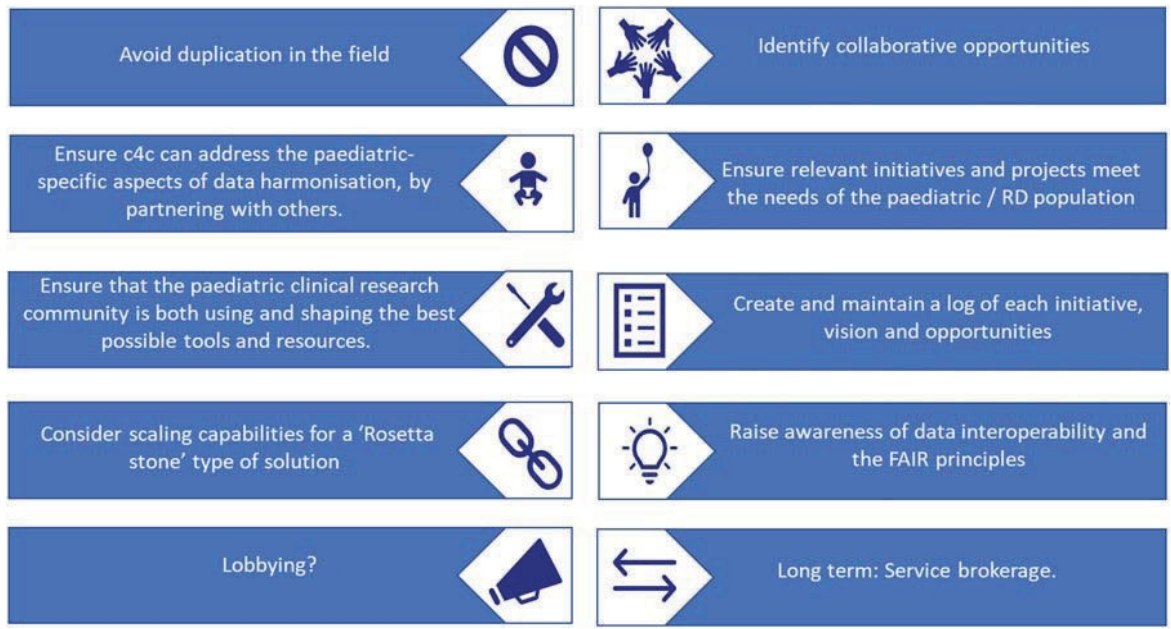

FIGURE 19.4 Goals and long-term vision for the GLOPAD.

- Ensure that the pediatric clinical research community is both using and shaping the best possible tools and resources.
- Consider scaling capabilities for a "Rosetta Stone" type of solution.
- Lobbying, for example pushing the EC on their understanding of standardization of data and what they mean by health data.
- Identify collaborative opportunities. Ensure relevant initiatives and projects meet the needs of the pediatric/RD population.
- Create and maintain a log of each initiative, vision, and opportunities.
- Raise awareness of data interoperability and the FAIR principles (FAIR metadata profile).

## Conclusions

As the chapter explains, many unmet needs remain, in relation to therapy development for pediatric rare diseases. A significant population is affected here, as an estimated 300 million people around the world live with a rare disease, and an estimated 70% of these will be pediatric patients. More innovative and supportive policies (at national, European, and indeed global level) are required, in order to foster research ecosystems in which research for PRD can prosper. It will be necessary to encourage a stronger pipeline of preclinical research, ideally addressing hitherto neglected conditions while improving therapeutic options for the better-studied diseasesdin order to make clinical trials a possibility for a broader range of pediatric conditions. Once trials are a reality, it will be necessary to scale-up application of the kinds of solutions described in this chapter, in order to overcome some of the traditional challenges around planning trials, finding experts, recruiting patients, and opening studies in multiple countries in a timely fashion. This will entail continued close collaboration between the rare disease and pediatric communities, while also leveraging expertise and advances in the broader arenas of clinical trials, data, PPI, and more. Moreover, as R&D and regulatory science evolve to embrace more innovative approaches to designing and delivering trials in small populations, and advanced therapy medicinal products such as gene and cell therapies continue to offer new hope to patients with rare genetic pediatric diseases; this community will need to place ever greater emphasis on learning from past endeavors, agreeing best practices, building capacity, and innovating on multiple levels, to allow the PRD community to deliver more clinical trials with better chances of success, ultimately leading to a larger number of effective treatments for this particularly vulnerable population.

## References

Aartsma-Rus, A., Dooms, M., & Le Cam, Y. (2021). Orphan medicine incentives: How to address the unmet needs of rare disease patients by optimizing the European orphan medicinal product landscape guiding principles and policy proposals by the European expert group for orphan drug incentives (OD expert group). *Frontiers in Pharmacology, 12*. https://doi.org/10.3389/fphar.2021.744532

Batchelor, H., Venables, R., Ranmal, S., & Tuleu, C. (2015). Oral formulations for paediatrics: Palatability studies. *Hospital Pharmacy Europe*. https://hospitalpharmacyeurope.com/news/editors-pick/oral-formulations-for-paediatrics-palatability-studies/.

Bladen, C. L., Rafferty, K., Straub, V., Monges, S., Moresco, A., Dawkins, H., Roy, A., Chamova, T., Guergueltcheva, V., Korngut, L., Campbell, C., Dai, Y., Barišić, N., Kos, T., Brabec, P., Rahbek, J., Lahdetie, J., Tuffery-Giraud, S., Claustres, M., ... Lochmüller, H. (2013). The TREAT-NMD duchenne muscular dystrophy registries: Conception, design, and utilization by industry and academia. *Human Mutation, 34*(11), 1449e1457. https://doi.org/10.1002/humu.22390

Bobbitt, D. R., LeRoy, B., & Palmer, a (2019). Clinical data standards and the new world of research science, technology, and data sources. *Journal of the National Institute of Public Health, 68*(3), 194e201. https://doi.org/10.20683/jniph.68.3_194

Boulanger, V., Schlemmer, M., Rossov, S., Seebald, A., & Gavin, P. (2020). Establishing patient registries for rare diseases: Rationale and challenges. *Pharmaceutical Medicine, 34*(3), 185e190. https://doi.org/10.1007/s40290-020-00332-1

Caldwell, P. H., Murphy, S. B., Butow, P. N., & Craig, J. C. (2004). Clinical trials in children. *Lancet, 364*(9436), 803e811. https://doi.org/10.1016/S0140-6736(04)16942-0

CDISC. (2021). *FHIR to CDISC joint mapping implementation guide v1.0* Accessed, 2, 2023 https://www.cdisc.org/standards/real-world-data/fhir-cdisc-joint-mapping-implementation-guide-v1-0.

Cheng, K., Mahler, F., Lutsar, I., Nafria Escalera, B., Breitenstein, S., Vassal, G., Claverol, J., Noel Palacio, N., Portman, R., Pope, G., Bakker, M., van der Geest, T., Turner, M. A., de Wildt, S. N., Lanzerath, D., Tuleu, C., Schwab, M., Smits, A., Treluyer, J.-M., … Cross, H. (2023). Clinical, methodology, and patient/carer expert advice in pediatric drug development by conect4children. *Clinical and Translational Science, 16*(3), 478e488. https://doi.org/10.1111/cts.13459

Cowie, M. R., Blomster, J. I., Curtis, L. H., Duclaux, S., Ford, I., Fritz, F., Goldman, S., Janmohamed, S., Kreuzer, J., Leenay, M., Michel, A., Ong, S., Pell, J. P., Southworth, M. R., Stough, W. G., Thoenes, M., Zannad, F., & Zalewski, A. (2017). Electronic health records to facilitate clinical research. *Clinical Research in Cardiology, 106*(1). https://doi.org/10.1007/s00392-016-1025-6

Davydov, A., Orlova, A., Didden, E.-M., Ong, R., Biedermann, P., Wetheril, G., & Klebanov, M. (2020). *Pathways for advanced transformation of CDISC SDTM data sets into OMOP CDM.* https://www.ohdsi.org/wp-content/uploads/2020/10/Michael-Kallfelz-Kallfelz_Pathways_SDTM-OMOP_abstract.pdf.

De Moor, G., Sundgren, M., Kalra, D., Schmidt, A., Dugas, M., Claerhout, B., Karakoyun, T., Ohmann, C., Lastic, P. Y., Ammour, N., Kush, R., Dupont, D., Cuggia, M., Daniel, C., Thienpont, G., & Coorevits, P. (2015). Using electronic health records for clinical research: The case of the EHR4CR project. *Journal of Biomedical Informatics, 53*, 162e173. https://doi.org/10.1016/j.jbi.2014.10.006

de Wildt, S. N., & Wong, I. C. K. (2022). Innovative methodologies in paediatric drug development: A conect4children (c4c) special issue. *British Journal of Clinical Pharmacology, 88*(12), 4962e4964. https://doi.org/10.1111/bcp.15355

Demotes-Mainard, J., & Kubiak, C. (2011). A European perspectiveeThe European clinical research infrastructures network. *Annals of Oncology, 22*(7), vii44evii49. https://doi.org/10.1093/annonc/mdr425

Desvignes-Gleizes, C., Vilcot, T., Donough, Mc, & et al.. (2022). *Facilitating and accelerating patient-reported outcomes (PRO) selection to use for clinical research and rare diseases (RD)-A European rare disease research coordination and support action (ERICA) project. Value in health* (p. 25).

Directorate-General For Food and Health Safety. (2022). *Recommendation paper on decentralised elements in clinical trials* Accessed, 2, 2023 https://health.ec.europa.eu/latest-updates/recommendation-paper-decentralised-elements-clinical-trials-2022-12-14_en.

dos Santos Vieira, B., Bernabe, C. H., Zhang, S., Abaza, H., Benis, N., Camara, A., Cornet, R., Le Cornec, C. M. A., t Hoen, P., Schaefer, F., Van der Velde, K. J., Swertz, M. A., Wilkinson, M. D., Jacobsen, A., & Ross, M. (2022). Towards FAIRification of sensitive and fragmented rare disease patient data: Challenges and solutions in European reference network registries. *Orphanet Journal of Rare Disease, 17*, Article 436. https://doi.org/10.1186/s13023-022-02558-5. In this issue.

Dupont, D., Beresniak, A., Kalra, D., Coorevits, P., & De Moor, G. (2018). Value of hospital electronic health records for clinical research: Contribution of the European project EHR4CR. *Medecine Sciences, 34*(11), 972e977. https://doi.org/10.1051/medsci/2018235

Eskola, S. M., Leufkens, H. G. M., Bate, A., De Bruin, M. L., & Gardarsdottir, H. (2022). Use of real-world data and evidence in drug development of medicinal products centrally authorized in Europe in 2018e2019. *Clinical Pharmacology and Therapeutics (St. Louis), 111*(1), 310e320. https://doi.org/10.1002/cpt.2462

Esserman, L. J., Berry, D. A., DeMichele, A., Carey, L., Davis, S. E., Buxton, M., Hudis, C., Gray, J. W., Perou, C., Yau, C., Livasy, C., Krontiras, H., Montgomery, L., Tripathy, D., Lehman, C., Liu, M. C., Olopade, O. I., Rugo, H. S., Carpenter, J. T., … Hylton, N. (2012). Pathologic complete response predicts recurrence-free survival more effectively by cancer subset: Results from the I-SPY 1 TRIALdCALGB 150007/150012, ACRIN 6657. *Journal of Clinical Oncology, 30*(26), 3242e3249. https://doi.org/10.1200/JCO.2011.39.2779

EU. (2000). Regulation (EC) No 141/2000 of the European Parliament and of the Council of 16 December 1999 on orphan medicinal products. *Official Journal of the European Union, 18*, 1e5.

EU. (2004). Regulation (EC) No 1901/2006 of the European Parliament and of the Council of 12 December 2006 on medicinal products for paediatric use and amending regulation (EEC) No 1768/92, directive 2001/20/EC, directive 2001/83/EC and regulation (EC) No 726/2004 (text with EEA relevance). *Official Journal of the European Union, 378*, 1e19.

EURODIS. (2020). *Improving patient access to gene and cell therapies for rare diseases in Europe* Accessed, 2, 2023 https://rareimpact.eu/site/wp-content/uploads/2020/04/RARE-IMPACT-Country-Assessments-Netherlands_v1_2020-04-28.pdf.

European Commission. European Reference Networks. https://webgate.ec.europa.eu/ernsd/cgi-bin/ern_public.cgi?npage ern_portal.html#!/. (Accessed 9 May 2024).

European Medicines Agency and its Paediatric Committee 2016 2017 commission report on the paediatric regulation. European Medicines Agency and its Paediatric Committee. https://ec.europa.eu/health/medicinal-products/medicines-children/2017-commission-report-paediatric-regulation_en. Accessed, 2, 2022.

Facile, R., Muhlbradt, E. E., Gong, M., Li, Q., Popat, V., Pětavy, F., Cornet, R., Ruan, Y., Koide, D., Saito, T. I., Hume, S., Rockhold, F., Bao, W., Dubman, S., & Jauregui Wurst, B. (2022). Use of clinical data interchange standards consortium (CDISC) standards for real-world data: Expert perspectives from a qualitative Delphi survey. *JMIR Medical Informatics, 10*(1). https://doi.org/10.2196/30363

FDA. (2020). *FDA guidance on conduct of clinical trials of medical products during the COVID-19 public health emergency.* https://www.fda.gov/regulatory-information/search-fda-guidance-documents/fda-guidance-conduct-clinical-trials-medical-products-during-covid-19-public-health-emergency Accessed, 0.

FDA. (2017). *Study data standard: What you need to know* Accessed, 2, 2022 https://www.fda.gov/files/drugs/published/Study-Data-Standards–What-You-Need-to-Know.pdf.

Felisi, M., Bonifazi, F., Toma, M., Pansieri, C., Hedley, V., Leary, R., Cornet, R., Reggiardo, R., Landi, A., D'Ercole, A., Malik, S., Nally, S., Sen, A., Palmeri, A., Bonifazi, D., & Ceci, A. (2024). Mapping of data-sharing repositories for paediatric clinical research–A rapid review. *Data, 9*(4), Article 59. https://doi.org/10.3390/data9040059. In this issue.

Freidlin, B., & Korn, E. L. (2017). Two-by-Two factorial cancer treatment trials: Is sufficient attention being paid to possible interactions? *Journal of the National Cancer Institute, 109*(9). https://doi.org/10.1093/jnci/djx146

Garcia, S., Bhatt, D. L., Gallagher, M., Jneid, H., Kaufman, J., Palevsky, P. M., Wu, H., & Weisbord, S. D. (2018). Strategies to reduce acute kidney injury and improve clinical outcomes following percutaneous coronary intervention: A subgroup analysis of the PRESERVE trial. *JACC: Cardiovascular Interventions, 11*(22), 2254e2261. https://doi.org/10.1016/j.jcin.2018.07.044

Gliklich, Dreyer, N. A., & Leavy, M. B. (2014). *Registries for evaluating patient outcomes: A user's guide.*

Goedecke, T., & Arlett, P. (2014). *A description of the European network of centres for Pharmacoepidemiology and pharmacovigilance as a global resource for pharmacovigilance and Pharmacoepidemiology* (pp. 403e408). Wiley. https://doi.org/10.1002/9781118820186.ch24

Gold, S. M., Bofill Roig, M., Miranda, J. J., Pariante, C., Posch, M., & Otte, C. (2022). Platform trials and the future of evaluating therapeutic behavioural interventions. *Nature Reviews Psychology, 1*(1), 7e8. https://doi.org/10.1038/s44159-021-00012-0

Hardi, R. (2010). *A four arm study to evaluate the safety and efficacy of 3 different doses of RVX-100 versus placebo in subjects with irritable bowel syndrome accompanied by diarrhea (IBS-D)* Accessed, 2, 2023 https://clinicaltrials.gov/ct2/show/NCT01076699.

Hedley, V., Kole, A., Rodwell, C., & Simon, F. (2019). *Rare 2030 knowledge base summary on data collection and utilisation for rare diseases*, 2019 Accessed, 2, 2023 http://download2.eurordis.org.s3.amazonaws.com/rare2030/Knowledge%20Based%20Summaries/Knowledge%20Base%20Summary%20-%20Data%20Collection%20and%20Utilisation%20%281%29.pdf.

Heslop, E., Turner, C., Irvin, A., Muntoni, F., Straub, V., & Guglieri, M. (2021). Gene therapy in Duchenne muscular dystrophy: Identifying and preparing for the challenges ahead. *Neuromuscular Disorders, 31*(1), 69e78. https://doi.org/10.1016/j.nmd.2020.10.001. Unpublished content. Hivert, V., Jonker, A. H., O'Connor, D., & Ardigo, D. (2021). IRDiRC: 1000 new rare diseases treatments by 2027, identifying and bringing forward strategic actions. *Rare Disease and Orphan Drugs Journal.* https://doi.org/10.20517/rdodj.2021.02

Horgan, D., Metspalu, A., Ouillade, M.-C., Athanasiou, D., Pasi, J., Adjali, O., Harrison, P., Hermans, C., Codacci-Pisanelli, G., Koeva, J., Szucs, T., Cursaru, V., Belina, I., Bernini, C., Zhuang, S., McMahon, S., Toncheva, D., & Thum, T. (2020). Propelling healthcare with advanced therapy medicinal products: A policy discussion. *Biomedicine Hub, 5*(3), 1e23. https://doi.org/10.1159/000511678

Jaki, T., & Wason, J. M. S. (2018). Multi-arm multi-stage trials can improve the efficiency of finding effective treatments for stroke: A case study. *BMC Cardiovascular Disorders, 18.* https://doi.org/10.1186/s12872-018-0956-4

James, N. D., De Bono, J. S., Spears, M. R., Clarke, N. W., Mason, M. D., Dearnaley, D. P., Ritchie, A. W. S., Amos, C. L., Gilson, C., Jones, R. J., Matheson, D., Millman, R., Attard, G., Chowdhury, S., Cross, W. R., Gillessen, S., Parker, C. C., Russell, J. M., Berthold, D. R., … Sydes, M. R. (2017). Abiraterone for prostate cancer not previously treated with hormone therapy. *New England Journal of Medicine, 377*(4), 338e351. https://doi.org/10.1056/NEJMoa1702900

James, N. D., Sydes, M. R., Clarke, N. W., Mason, M. D., Dearnaley, D. P., Anderson, J., Popert, R. J., Sanders, K., Morgan, R. C., Stansfeld, J., Dwyer, J., Masters, J., & Parmar, M. K. B. (2009). Systemic therapy for advancing or metastatic prostate cancer (STAMPEDE): A multi-arm, multistage randomized controlled trial. *BJU International, 103*(4), 464e469. https://doi.org/10.1111/j.1464-410X.2008.08034.x

Julkowska, D., Austin, C. P., Cutillo, C. M., Gancberg, D., Hager, C., Halftermeyer, J., Jonker, A. H., Lau, L. P. L., Norstedt, I., Rath, A., Schuster, R., Simelyte, E., & Van Weely, S. (2017). The importance of international collaboration for rare diseases research: A European perspective. *Gene Therapy, 24*(9), 562e571. https://doi.org/10.1038/gt.2017.29

Kole, A., & Hedley, V. (2021). *Recommendations from the rare 2030 foresight study. The future of rare diseases starts today.* https://download2.eurordis.org/rare2030/Rare2030_recommendations.pdf.

Kole, A., & Faurisson, F. (2009). *The voice of 12,000 patients. Experiences and expectations of rare disease patients on diagnosis and care in Europe.* EURORDIS-Rare Diseases EU.

Kole, A. (2019). *Validation of trends to build future scenarios in rare disease policy.* https://download2.eurordis.org/rare2030/deliverables/Rare%202030%20Panel%20of%20Experts%20Conference%20Report.pdf.

Leary, R., Oyewole, A. O., Bushby, K., & Aartsma-Rus, A. (2017). Translational research in Europe for the assessment and treatment for neuromuscular disorders (TREAT-NMD). *Neuropediatrics, 48*(4), 211e220. https://doi.org/10.1055/s-0037-1604110

Lepola, P., Needham, A., Mendum, J., Sallabank, P., Neubauer, D., & de Wildt, S. (2016). Informed consent for paediatric clinical trials in Europe. *Archives of Disease in Childhood, 101*(11), 1017e1025. https://doi.org/10.1136/archdischild-2015-310001

Lim, A., & Cranswick, N. (2003). Clinical trials, ethical issues and patient recruitment: An Australian perspective. *Paediatric and Perinatal Drug Therapy, 5*(4), 183e187. https://doi.org/10.1185/146300903774115775

Liu, W., Zhou, L., Feng, L., Zhang, D., Zhang, C., & Gao, Y. (2021). BuqiTongluo granule for ischemic stroke, stable angina pectoris, diabetic peripheral neuropathy with Qi deficiency and blood stasis syndrome: Rationale and novel basket design. *Frontiers in Pharmacology, 12.* https://doi.org/10.3389/fphar.2021.764669

McMullan, J., Crowe, A. L., McClenaghan, T., McAneney, H., & McKnight, A. J. (2021). Perceptions and experiences of rare

diseases among general practitioners: An exploratory study. *medRxiv*. medRxiv. https://doi.org/10.1101/2021.09.07.21263025

McNair, L. (2020). Basket clinical trial designs: The key to testing innovative therapies is innovation in study design and conduct. *The Clinical Researcher, 34*(2).

Mills, E. J., Chan, A. W., Wu, P., Vail, A., Guyatt, G. H., & Altman, D. G. (2009). Design, analysis, and presentation of crossover trials. *Trials, 10*. https://doi.org/10.1186/1745-6215-10-27

Mistry, P. K., Kishnani, P., Wanner, C., Dong, D., Bender, J., Batista, J. L., & Foster, J. (2022). Rare lysosomal disease registries: Lessons learned over three decades of real-world evidence. *Orphanet Journal of Rare Diseases, 17*(1). https://doi.org/10.1186/s13023-022-02517-0

Montgomery, A. A., Peters, T. J., & Little, P. (2003). Design, analysis and presentation of factorial randomised controlled trials. *BMC Medical Research Methodology, 3*, 1e5. https://doi.org/10.1186/1471-2288-3-26

Morrow, D., Ussi, A., & Migliaccio, G. (2017). Addressing pressing needs in the development of advanced therapies. *Frontiers in Bioengineering and Biotechnology, 5*. https://doi.org/10.3389/fbioe.2017.00055

Mulleman, D., Mammou, S., Griffoul, I., Watier, H., & Goupille, P. (2006). Pathophysiology of disk-related sciatica. I.dEvidence supporting a chemical component. *Joint Bone Spine, 73*(2), 151e158. https://doi.org/10.10 16/j.jbspin.2005.03.003

Newton, W. (2021). Basket and umbrella trials: Untapped op- portunities in rare disease. *Clinical Trials Arena*. https://www.clinicaltrialsarena.com/features/basket- and- umbrella-trials-untapped-opportunities-in-rare-dis- ease/.

Odutayo, A., Gryaznov, D., Copsey, B., Monk, P., Speich, B., Roberts, C., Vadher, K., Dutton, P., Briel, M., Hopewell, S., & Altman, D. G. (2020). Design, analysis and reporting of multi-arm trials and strategies to address multiple testing. *International Journal of Epidemiology, 49*(3), 968e978. https://doi.org/10.1093/ije/dyaa026

Palmeri, A., Leary, R., Sen, A., Cornet, R., Welter, D., & Rocca-Serra, P. (2023). Creating a metadata profile for clinical trial protocols. https://faircookbook.elixir-europe.org/content/recipes/interoperability/c4c-clinical-trials.html. (Accessed 8 May 2023).

Pericàs, J. M., Derde, L. P. G., & Berry, S. M. (2023). Platform trials as the way forward in infectious disease' clinical research: The case of coronavirus disease 2019. *Clinical Microbiology and Infection, 29*(3), 277e280. https:// doi.org/10.1016/j.cmi.2022.11.022

Pericàs, J. M., Tacke, F., Anstee, Q. M., Di Prospero, N. A., Kjær, M. S., Mesenbrink, P., Koenig, F., Genescà, J., & Ratziu, V. (2023). Platform trials to overcome major short-comings of traditional clinical trials in non-alcoholic stea- tohepatitis? Pros and cons. *Journal of Hepatology, 78*(2), 442e447. https://doi.org/10.1016/j.jhep.2022.09.021

Pletcher, M. J., Flaherman, V., Najafi, N., Patel, S., Rushakoff, R. J., Hoffman, A., Robinson, A., Cucina, R. J., McCulloch, C. E., Gonzales, R., & Auerbach, A. (2020). Randomized controlled trials of electronic health record interventions: Design, conduct, and reporting considerations. *Annals of Internal Medicine, 172*(11_Suppl.), S85eS91. https://doi.org/ 10.7326/m19-0877

Preston, J., Stones, S. R., Davies, H., Preston, J., & Phillips, B. (2019). How to involve children and young people in what is, after all, their research. *Archives of Disease in*

*Childhood, 104*(5), 494e500. https://doi.org/10.1136/archdischild-2018-315118

*Rare disease Moonshot.*(2023) Accessed, 2, 2023 https://www.rarediseasemoonshot.eu/.

Reed, R. C., Rosenfeld, W. E., Lippmann, S. M., Eijkemans, R. M. J. C., & Kasteleijn-Nolst Trenité, D. G. A. (2020). Rapidity of CNS effect on photo- paroxysmal response for brivaracetam vs. Levetiracetam: A randomized, double-blind, crossover trial in photosensitive epilepsy patients. *CNS Drugs, 34*(10), 1075e1086. https://doi.org/10.1007/s40263-020-00761-1

Robert, C., Ribas, A., Wolchok, J. D., Hodi, F. S., Hamid, O., Kefford, R., Weber, J. S., Joshua, A. M., Hwu, W. J., Gangadhar, T. C., Patnaik, A., Dronca, R., Zarour, H., Joseph, R. W., Boasberg, P., Chmielowski, B., Mateus, C., Postow, M. A., Gergich, K., … Daud, A. (2014). Anti-programmed-death-receptor-1 treatment with pembrolizumab in ipilimumab-refractory advanced melanoma: A randomised dose-comparison cohort of a phase 1 trial. *The Lancet, 384*(9948), 1109e1117. https://doi.org/10.1016/S0140-6736(14)60958-2

Rode, J. (2005). *Rare diseases: Understanding this public health priority* (p. 5). EURORDIS. https://www.myeloma.ro/documente/european-organization-for-rare-diseases-november-2005.pdf.

Rodger, S., Lochmüller, H., Tassoni, A., Gramsch, K., König, K., Bushby, K., Straub, V., Korinthenberg, R., & Kirschner, J. (2013). The TREAT-NMD care and trial site registry: An online registry to facilitate clinical research for neuromuscular diseases. *Orphanet Journal of Rare Diseases, 8*(1). https://doi.org/10.1186/1750-1172-8-171

Roig, M. B., Krotka, P., Burman, C. F., Glimm, E., Gold, S. M., Hees, K., Jacko, P., Koenig, F., Magirr, D., Mesenbrink, P., Viele, K., & Posch, M. (2022). On model-based time trend adjustments in platform trials with non-concurrent controls. *BMC Medical Research Methodology, 22*(1). https://doi.org/10.1186/s12874-022-01683-w

Rosterman, J. L., Pallotto, E. K., Truog, W. E., Escobar, H., Meinert, K. A., Holmes, A., Dai, H., & Manimtim, W. M. (2018). The impact of neurally adjusted ventilatory assist mode on respiratory severity score and energy expenditure in infants: A randomized crossover trial. *Journal of Perinatology, 38*(1), 59e63. https://doi.org/10.1038/jp.2017.154

Schmidli, H., Häring, D. A., Thomas, M., Cassidy, A., Weber, S., & Bretz, F. (2020). Beyond randomized clinical trials: Use of external controls. *Clinical Pharmacology and Therapeutics (St. Louis), 107*(4), 806e816. https://doi.org/10.1002/cpt.1723

Sen, A., Hedley, V., Degraeuwe, E., Hirschfeld, S., Cornet, R., Walls, R., Owen, J., Robinson, P. N., Neilan, E. G., Leiner, T., Nisato, G., Modi, N., Woodworth, S., Palmeri, A., Gaentzsch, R., Walsh, M., Berkery, T., Lee, J., Persijn, L., Baker, K., An Haack, K., Segovia Simon, S., Jacobsen, J. O. B., Reggiardo, G., Kirwin, M. A., Trueman, J., Pansieri, C., Bonifazi, D., Nally, S., Bonifazi, F., Leary, R., & Straub, V. (2024). Learning from conect4children: A collaborative approach towards Standardisation of Disease-Specific Paediatric Research Data. *Data, 9*(4), Article 55. https://doi.org/10.3390/data9040055. In this issue.

Sen, A., Hedley, V., Owen, J., Cornet, R., Kalra, D., Engel, C., Palmeri, A., Lee, J., Roze, J. C., Standing, J. F., Warris, A., Pansieri, C., Leary, R., Turner, M., & Straub, V. (2023). Standardizing paediatric clinical data: The development of the conect4children (c4c) cross cutting paediatric data dictionary. *Journal of the Society for Clinical Data Management, 2*(3). https://doi.org/10.47912/jscdm.218. In this issue.

Sen, A., Palmeri, A., Lee, J., Hedley, V., Thuet, J., Lignon, P., Cotonnec, V., Leary, R., Nally, S., & Straub, V. (2023). Understanding paediatric data standards challenges through academia-industry partnerships: A conect4children (c4c) qualitative study. *The International Journal of Health Planning and Management, 38*(2), 416e429. https://doi.org/10.1002/hpm.3592

Simon, K. C., Tideman, S., Hillman, L., Lai, R., Jathar, R., Ji, Y., Bergman-Bock, S., Castle, J., Franada, T., Freedom, T., Marcus, R., Mark, A., Meyers, S., Rubin, S., Semenov, I., Yucus, C., Pham, A., Garduno, L., Szela, M., Frigerio, R., & Maraganore, D. M. (2018). Design and implementation of pragmatic clinical trials using the electronic medical record and an adaptive design. *JAMIA Open, 1*(1), 99e106. https://doi.org/10.1093/jamiaopen/ooy017

Staffa, J. A., & Dal Pan, G. J. (2012). Regulatory innovation in postmarketing risk assessment and management. *Clinical Pharmacology and Therapeutics (St. Louis), 91*(3), 555e557. https://doi.org/10.1038/clpt.2011.289

Stirnadel-Farrant, H., Kudari, M., Garman, N., Imrie, J., Chopra, B., Giannelli, S., Gabaldo, M., Corti, A., Zancan, S., Aiuti, A., Cicalese, M. P., Batta, R., Appleby, J., Davinelli, M., & Ng, P. (2018). Gene therapy in rare diseases: The benefits and challenges of developing a patient-centric registry for strimvelis in ADA-SCID. *Orphanet Journal of Rare Diseases, 13*(1). https://doi.org/10.1186/s13023-018-0791-9

Tambuyzer, E., Vandendriessche, B., Austin, C. P., Brooks, P. J., Larsson, K., Miller Needleman, K. I., Valentine, J., Davies, K., Groft, S. C., Preti, R., Oprea, T. I., & Prunotto, M. (2020). Therapies for rare diseases: Therapeutic modalities, progress and challenges ahead. *Nature Reviews Drug Discovery, 19*(2), 93e111. https://doi.org/10.1038/s41573-019-0049-9

Taruscio, D., Capozzoli, F., & Frank, C. (2011). Rare diseases and orphan drugs. *Annali dell'Istituto Superiore di Sanita, 47*(1), 83e93. https://doi.org/10.4415/ANN-11-01-17

Taruscio, D., Gainotti, S., Mollo, E., Vittozzi, L., Bianchi, F., Ensini, M., & Posada, M. (2013). The current situation and needs of rare disease registries in Europe. *Public Health Genomics, 16*(6), 288e298. https://doi.org/10.1159/000355934

Taruscio, D., Vittozzi, L., Choquet, R., Heimdal, K., Iskrov, G., Kodra, Y., Landais, P., Posada, M., Stefanov, R., Steinmueller, C., Swinnen, E., & Van Oyen, H. (2015). National registries of rare diseases in Europe: An overview of the current situation and experiences. *Public Health Genomics, 18*(1), 20e25. https://doi.org/10.1159/000365897

Turner, M. A., Hildebrand, H., Fernandes, R. M., de Wildt, S. N., Mahler, F., Hankard, R., Leary, R., Bonifazi, F., Nobels, P., Cheng, K., Attar, S., Rossi, P., Rocchi, F., Claverol, J., Nafria, B., & Giaquinto, C. (2021). The conect4children (c4c) consortium: Potential for improving European clinical research into medicines for children. *Pharmaceutical Medicine, 35*(2), 71e79. https://doi.org/10.1007/s40290-020-00373-6

Turner, M. A., Cheng, K., de Wildt, S., Hildebrand, H., Attar, S., Rossi, P., Bonifazi, D., Ceci, A., Claverol, J., Nafria, B., & Giaquinto, C. (2022). European research networks to facilitate drug research in children. *British Journal of Clinical Pharmacology, 88*(10), 4258e4266. https://doi.org/10.1111/bcp.14545

Turner, N. C., Kingston, B., Kilburn, L. S., Kernaghan, S., Wardley, A. M., Macpherson, I. R., Baird, R. D., Roylance, R., Stephens, P., Oikonomidou, O., Braybrooke, J. P., Tuthill, M., Abraham, J., Winter, M. C., Bye, H., Hubank, M., Gevensleben, H., Cutts, R., Snowdon, C., … Ring, A. (2020). Circulating tumour DNA analysis to direct therapy in advanced breast cancer (plasmaMATCH): A multicentre, multicohort, phase 2a, platform trial. *The Lancet Oncology, 21*(10), 1296e1308. https://doi.org/10.1016/S1470-2045(20)30444-7

Walkowiak, D., & Domaradzki, J. (2021). Are rare diseases overlooked by medical education? Awareness of rare diseases among physicians in Poland: An explanatory study. *Orphanet Journal of Rare Diseases, 16*(1). https://doi.org/10.1186/s13023-021-02023-9

Yoon, P. W., Olney, R. S., Khoury, M. J., Sappenfield, W. M., Chavez, G. F., & Taylor, D. (1997). Contribution of birth defects and genetic diseases to pediatric hospitalizations: A population-based study. *Archives of Pediatrics and Adolescent Medicine, 151*(11), 1096e1103. https://doi.org/10.1001/archpedi.1997.02170480026004

Zhu, Q., Nguyễn, Đ. T., Sheils, T., Alyea, G., Sid, E., Xu, Y., Dickens, J., Mathè, E. A., & Pariser, A. (2021). Scientific evidence based rare disease research discovery with research funding data in knowledge graph. *Orphanet Journal of Rare Diseases, 16*(1). https://doi.org/10.1186/s13023-021-02120-9